# A mechanistic verification of the competitive exclusion principle

Lev V. Kalmykov[1,3] & Vyacheslav L. Kalmykov[2,3*]

[1]Institute of Theoretical and Experimental Biophysics, Russian Academy of Sciences, Pushchino, Moscow Region, 142290 Russia; [2]Institute of Cell Biophysics, Russian Academy of Sciences, Pushchino, Moscow Region, 142290 Russia; [3]Pushchino State Institute of Natural Sciences (the former Pushchino State University), Pushchino, Moscow Region, 142290 Russia.

*e-mail: vyacheslav.l.kalmykov@gmail.com

## Abstract

Biodiversity conservation becoming increasingly urgent. It is important to find mechanisms of competitive coexistence of species with different fitness in especially difficult circumstances - on one limiting resource, in isolated stable uniform habitat, without any trade-offs and cooperative interactions. Here we show a mechanism of competitive coexistence based on a soliton-like behaviour of population waves. We have modelled it by the logical axiomatic deterministic individual-based cellular automata method. Our mechanistic models of population and ecosystem dynamics are of white-box type and so they provide direct insight into mechanisms under study. The mechanism provides indefinite coexistence of two, three and four competing species. This mechanism violates the known formulations of the competitive exclusion principle. As a consequence, we have proposed a fully mechanistic and most stringent formulation of the principle.

## Introduction

**Background.** Up to now, it is not clear why are there so many superficially similar species existing together[1,2]. The competitive exclusion principle (also known as Gause’s principle, Gause’s Rule, Gause’s Law, Gause’s Hypothesis, Volterra-Gause Principle, Grinnell’s Axiom, and Volterra-Lotka Law)[3] postulates that species competing for the same limiting resource in one homogeneous habitat cannot coexist[4,5]. This principle contradicts with the observed biodiversity. This contradiction is known as the biodiversity paradox[6]. Because of the long standing biodiversity debates[1,2,4,5], this paradox is the central problem of theoretical ecology. At least 120 different explanations of the biodiversity paradox were presented[3], however the biodiversity debates still continue[1,2]. Different competitive trade-offs[7-11] do not violate the competitive exclusion principle. The point is that the classic study of the principle in the work of Gause necessarily suggested that "one of species has any advantage over the other"[4], i.e. one and the same species *always* keeps a definite *uncompensated* benefit.

Fundamentally important point is that empirical studies cannot prove and hardly falsify the competitive exclusion principle – it can be verified only theoretically[5]. However, the theoretical studies of the subject left too many unresolved questions, what is one of the reasons of continuing biodiversity debates. Let us to consider the initial source of the problems with theoretical understanding of natural biodiversity. The Lotka-Volterra model predicts coexistence of two species when, for both species, an interspecific competition is weaker than intraspecific one. This interpretation follows directly from the Lotka-Volterra model. However, the further interpretation of this interpretation, known as *the competitive exclusion principle* has no rigorous justification under itself. The problem is that the Lotka-Volterra model is phenomenological[12]. Therefore this model cannot show underlined mechanisms of ecosystem under study in details. The Unified Neutral Theory of Biodiversity was proposed as an alternative to the niche theory. But Hubbell's neutral model is also discussed as "just a statement of ignorance about which species can succeed and why"[2]. The general problem is the absence of a mechanistic understanding of how and why many superficially similar species existing together.

In this paper we investigate a mechanism of competitive coexistence of species with different fitness in especially difficult circumstances - on one limiting resource, without any trade-offs and cooperative interactions, in isolated stable uniform habitat, and so on. We model a mechanistic mechanism of competitive coexistence outside the limitations imposed by the known formulations of the competitive exclusion principle.

**Main hypothesis.** We suppose that presence of free resource gaps may be a basis for competitive coexistence and for violation of the known formulations of the competitive exclusion principle. Free resource gaps may help to eliminate direct conflicts of interest between competing species. As result, colliding population waves of different competing species may interpenetrate through each other like soliton waves in physical systems. A possible mechanism of appearance of such gaps is a moderate reproduction.

**On a soliton-like behaviour of population waves.** Population waves are self-sustaining waves which use resources of a medium where they propagate. Population waves are autowaves[13]. Autowaves play a universal role in mechanisms of various chemical and biological processes[14-16]. The importance of autowaves is based on the universality of their properties that are independent of a specific implementation. One such universal property is that identical autowaves annihilate each other after collision. Previously, the paradoxical phenomenon of soliton-like (quasi-soliton) behaviour of population waves was revealed for ultra-fast chemotactic bacteria - their colliding population waves did not annihilate each other and looked as penetrating through or reflecting from each other without significant delay[13,17,18]. The phenomenon of soliton-like behaviour of chemotactic bacterial waves is based on ultrafast bacterial movement and, as the result, bacteria have no time to use all local resource[13,17]. Thus, a certain amount of unused resource is left behind the population wave front. Consequently, a reflection of the chemotactic waves and, possibly, their interpenetration through each other may occur after their collision. The question remains - what is implemented exactly - reflection, interpenetration, or both. Individual bacteria of colliding population waves were not marked[13,17,18]. So, a mechanistic mechanism of soliton-like interpenetration of colliding population waves was not directly demonstrated. Modelling by differential equations[19] has not helped to realize what really happens because it does not show what happens with population waves on a micro-level of individuals and their local interactions. Moreover the results on chemotactic soliton-like behaviour were obtained on one and the same species while we look for a mechanism of interspecific competitive coexistence.

A similar task of distinguishing between interpenetration and reflection arose in the studies of colliding population waves of the bacteria Myxococcus Xanthus[20-22]. Under starvation conditions these bacteria start to act cooperatively, aggregate and finally build a multicellular structure, the fruiting body. The fruiting body formation is often preceded by the pattern of periodically colliding waves called rippling patterns. In the difference from chemotactic bacteria, myxobacterial aggregation is the consequence of direct cell-to-cell contact interactions, but not of chemoreception of a food concentration gradient. When viewed from a distance where only cell density can be perceived, the rippling waves appear to pass through one another, analogous to soliton waves in various physical systems[22,23]. Nevertheless, the detailed studies of the population waves' behaviour of myxobacteria showed that they actually reflect off one another when they collide. Each wave crest oscillates back and forth with no net displacement. Without observing individual cells, the illusion that the waves pass through one another is nearly perfect. In the experimental study of colliding population waves some individual cells were marked by green fluorescent protein[23,24] and in computational modelling the agent-based approach was implemented[23,25]. These results ruled out the assumption about the soliton-like behaviour of rippling population waves of myxobacteria.

**On a methodology.** Phenomenological models are difficult to interpret mechanistically and thus they are of limited usefulness[12]. They may superficially show what happens with modelled object but cannot show how it happens in physical sense. They describe some empirical observations, but have no foundations in mechanisms or first principles. Phenomenological models are rather of 'black-box' type what means that they are not transparent for understanding of underlying mechanisms.

In contrast to the phenomenological black-box models, cellular automata models can be of a '*white box*' type. A white box model has 'transparent walls' and directly shows underlined mechanistic mechanisms – all events at micro-, mini- (meso-) and macro- levels of the simulated dynamic system are directly visible at all stages. We consider *mechanistic mechanisms* as consisting of two inter-related constituents – (1) cause-effect relations and (2) part-whole relations. The causes of the mechanisms should be sufficient to understand their effects and the parts should be sufficient to understand the whole. In our research, the 'whole' is a specific ecosystem with populations of competing species. The 'parts' are individuals, microhabitats and interactions between the individuals and their immediate environment (neighbourhood).

Mechanicalness of models may be achieved by using individual-based cellular automata approach. A logical deterministic individual-based cellular automata modelling may help us to find a mechanism of coexistence by the simplest way. The methodology of our approach has been described in detail previously[26]. Currently, the white-box modelling is mostly used in engineering sciences[27-29].

## Results

The main goal of this paper is a mechanistic verification of the competitive exclusion principle. To achieve that we have elaborated a cellular automata model of interspecific competition. Cellular automata provide mechanistic simulations of spatial and temporal relationships between individuals of competing species. This modelling method is based on physically interpreted ecological axioms i.e. on first principles (Fig. S2). The cellular automata model works as the automaton of visualised inference of a mechanistic insight into studied phenomena (Movies S1-S4).

**A mechanism of competitive coexistence.** On the basis of our main hypothesis we have proposed a special cellular automata neighbourhood with gaps (Fig. 5a). This spatial hexagonal rosette-like pattern of the offsprings placement permit to model population waves with free resource gaps. In accordance with the neighbourhood, offsprings of every individual can occupy no more than one third of resources of nearest environment of their parental plant. Rosettes of rhizomes of asexually propagating turf grasses *Poa pratensis* L. and *Festuca rubra* L. *ssp. rubra* are biological prototypes of the specific form of this neighbourhood.

We have found that moderate reproduction of competing species may be a base for soliton-like behaviour of colliding population waves (Fig. 1; Movie S1). Individuals of the colliding population waves of competing species freely interpenetrate through arising gaps. Therefore, interpenetration of colliding population waves occurs as tunnelling (interdigitation) of offsprings through the gaps in population waves (Fig. 1a; Movie S1). As a result, both competing species coexist indefinitely with the same equal numbers of individuals (Fig. 1b).

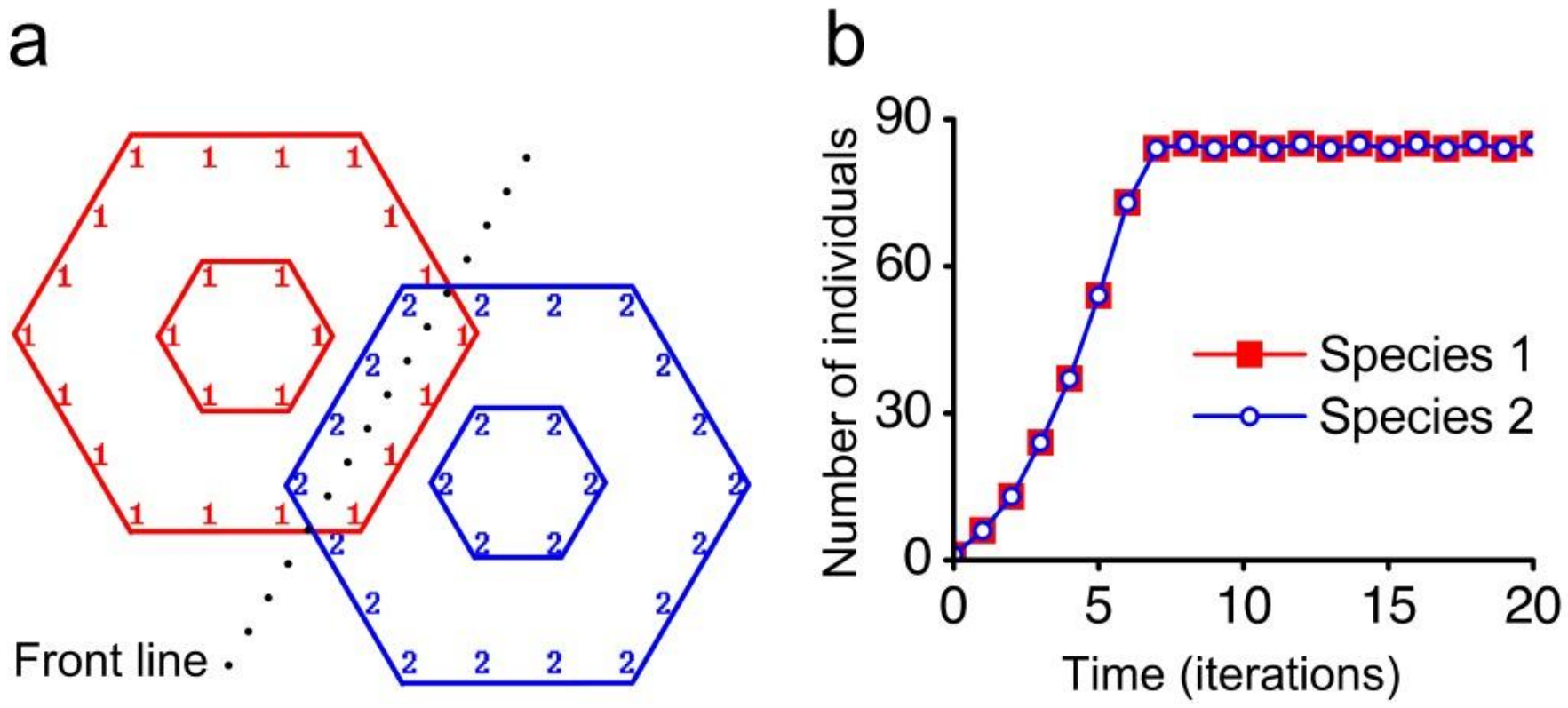

**Figure 1 | A soliton-like interpenetration of colliding population waves. a**, Outline of the interpenetration of colliding fronts of hexagonal population waves at the third iteration of the model (Movie S1). '1' - an individual of the first (dominant) species. '2' - an individual of the second (recessive) species. The dotted line represents a front line of the colliding population waves. **b**, Population dynamics of the two competing species in result of the free interpenetration of colliding population waves.

Additionally we show indefinite competitive coexistence of three and four species in one closed homogeneous habitat (Movies S3 and S4). All these species are complete competitors. They compete for one limiting resource in isolated stable uniform habitat without any trade-offs and cooperative interactions. A mechanism of their coexistence is based on moderate reproduction of individuals of all competing species. Unusually large numbers of cases when the less fit species survive have been found in the models of competition between two, three and four species (Fig. S5). In addition, we have shown that the species can coexist with each other in all possible combinations (Figs S6 and S7). Along with the cases of competitive coexistence (Movies S1, S3 and S4) we have shown also a case of competitive exclusion when competitors cannot avoid direct conflicts of interest (Movie S2).

**Investigation of population dynamics.** We use Monte Carlo simulations to investigate population dynamics of two, three and four competing species (Figs 2, S3-S8). The Monte Carlo simulation consists of multiple repeated experiments in which the initial conditions of the models are randomly changed. These multiple repeated experiments consist of random placements of single initial individuals of each species on the lattice. We show that competitive coexistence depends on initial positioning of individuals on the lattice (Fig. 2). Along with initial positioning of individuals on the lattice, the outcome of competition also depends on the lattice size. Coexistence occurs if the lattice consists of NxN lattice sites, where N is an even number (Figs 2, S3 and S4). If N is an odd number then with this type of the neighbourhood (Fig. 5a) the coexistence mechanism cannot be implemented and species cannot coexist (Figs S3 and S4).

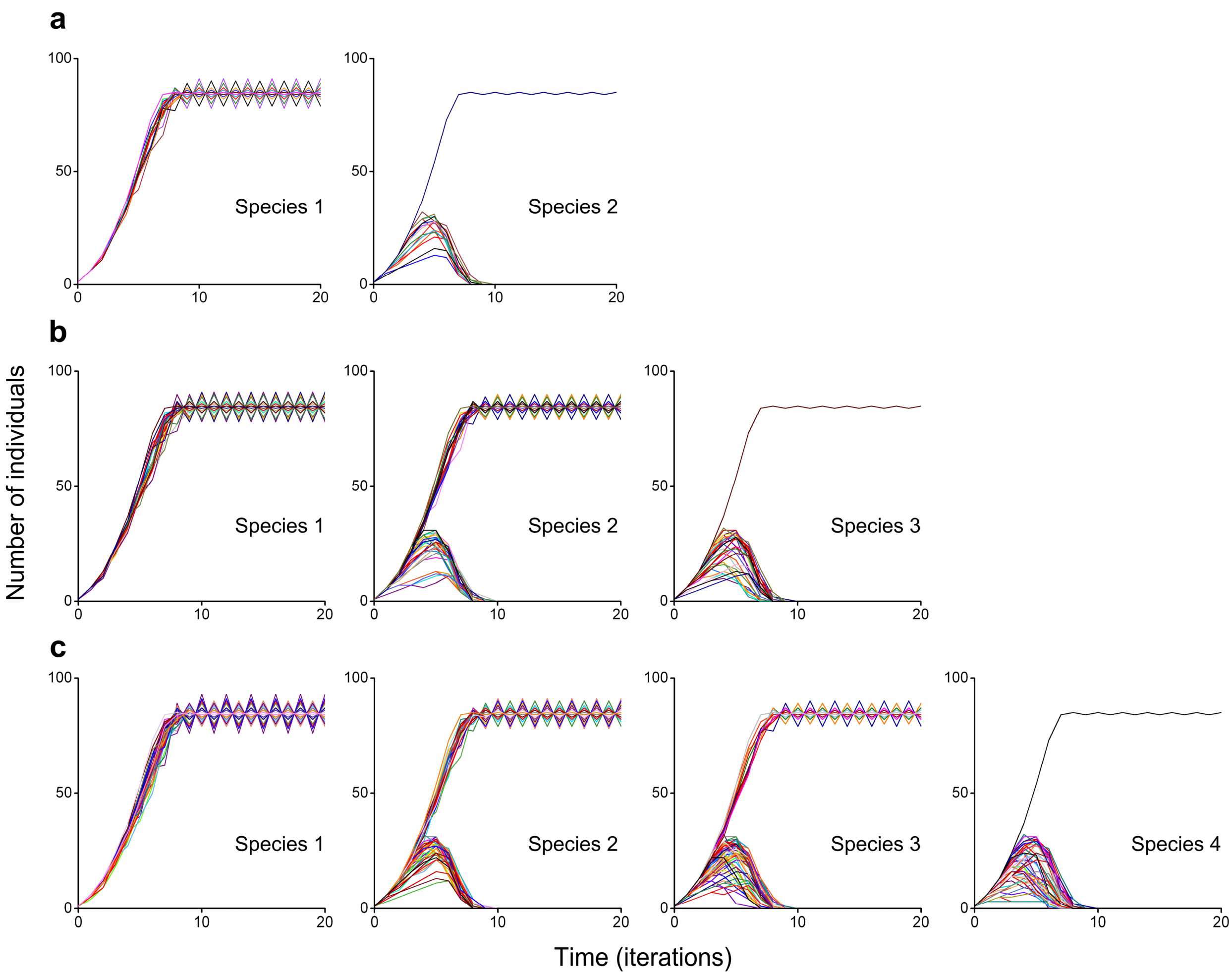

**Figure 2 | Population dynamics in the models with two, three and four competitors.** Random initial placements of single individuals of competing species on the lattice were used in each of 200 repeated experiments (Monte Carlo simulation). The lattice consists of 26x26 sites. **a**, Experiments with the two-species competition model. **b**, Experiments with the three-species competition model. **c**, Experiments with the four-species competition model.

## Discussion

The mechanism of competitive coexistence investigated here is a special case of spatio-temporal resource partitioning in isolated stable uniform habitat. It based on a soliton-like interpenetration of colliding population waves of competing species and implements the optimal allocation of a limiting resource among competitors enabling them to avoid direct conflicts and to maintain equal numbers of individuals in populations. This mechanism of indefinite competitive coexistence is based on a specific combination of the four basic conditions:

1. Moderate reproduction of individuals of competing species (number of offsprings is determined by the hexagonal rosette-like cellular automata neighbourhood);
2. Geometrically defined spatial placement of offsprings (it is determined by the hexagonal rosette-like cellular automata neighbourhood);
3. The definite structure and the definite size of an ecosystem (they are determined by the cellular automata lattice);
4. The definite initial spatial placement of single individuals of competing species (determined by their coordinates).

A rigorous theoretical verification of a validity of the known formulations of the principle may allow to solve the theoretical problem of contradictions between the principle and the observed biodiversity. In this paper we look for a new mechanism of competitive coexistence which contradicts the known formulations of the competitive exclusion principle. Such mechanism has been found. It provides indefinite coexistence of competing species, which are identical consumers and compete for one limiting resource in one isolated stable and uniform habitat. Here are the main conditions of our models:

1. There are no any trade-offs and cooperative interactions between the competing species;
2. Reproduction of the competing species occurs only vegetatively and the species are genetically homogenous and stable;
3. Individuals of one and the same species always win individuals of other competing species in direct conflict of interest;
4. A habitat is limited, homogeneous, stable, closed for immigration, emigration, predation, herbivory, parasitism and other disturbances;
5. Competing species are per capita identical and constant in ontogeny, in fecundity rates, in regeneration features of a microhabitat and in environmental requirements.

Our models completely satisfy to all these conditions (Figs 3-5, S1 and S2; Movies S1-S4). The known hypotheses of natural richness are based on non-compliance with one or more of these conditions[3,30,31]. Let us cite three of the most known formulations of the competitive exclusion principle:

"*n species require at least n resources*"[32];

"*no stable equilibrium can be attained in an ecological community in which some r of the components are limited by less than r limiting factors*"[33];

"*Two populations (species) cannot long coexist if they compete for a vital resource limitation of which is the direct and only factor limiting both populations. As thus restated, the principle is, I think, valid without exception*"[34].

Palmer carefully analyzed twelve formulations of the competitive exclusion principle and offered his own reformulated principle[3]:

"*Given a suite of species, interspecific competition will result in the exclusion of all but one species.*

*Conditions of the Principle:*

*(1) Time has been sufficient to allow exclusion;*
*(2) The environment is temporally constant;*
*(3) The environment has no spatial variation;*
*(4) Growth is limited by one resource;*
*(5) Rarer species are not disproportionately favored in terms of survivorship, reproduction, or growth;*
*(6) Species have the opportunity to compete;*
(7) *There is no immigration.*

*Corollary: The greater the degree to which these conditions are broken, the greater the number of species which can coexist.*"

Our models violate the wordings of Darlington[34] and Palmer[3] by showing the mechanism of competitive coexistence on one limiting resource. Furthermore, we have revealed a rather unexpected result that increasing the number of similar competing species led to the increased number of coexistence cases what contradicts to the cited formulations (Figs 2, S6-S8). Since our results violate the listed and other known formulations of the competitive exclusion principle, this principle should be additionally reformulated. We reformulate the competitive exclusion principle as follows:

> *If each and every individual of a less fit species in any attempt to use any limiting resource always has a direct conflict of interest with an individual of a most fittest species and always loses, then, all other things being equal for all individuals of the competing species, these species cannot coexist indefinitely and the less fit species will be excluded from the habitat in the long run.*

Our formulation of the principle is fully mechanistic. Implementation of this extremely strict formulation of the competitive exclusion principle is rather a very rare case in nature. This fact eliminates old contradictions between the competitive exclusion principle and natural biodiversity.

## Methods

A system of logical rules of transitions between the states of a lattice site of the cellular automata is formulated as a part of axioms on the basis of which we model population and ecosystem dynamics. The entire cellular automaton simulates a whole ecosystem which autonomously maintains and regenerates itself. A two-dimensional hexagonal lattice is closed to a torus by periodic boundary conditions in order to avoid boundary effects. We use the hexagonal lattice because it most naturally implements the principle of densest packing of microhabitats. Each site of the lattice simulates a microhabitat. In the free state a microhabitat contains resources for existence of a one individual of any species. An individual can occupy a one microhabitat only. A life cycle of an individual lasts a one iteration of the automaton. All states of all sites have the same duration. Every individual of all species consumes identical quantity of identical resources by identical way i.e. they are identical per capita consumers. Such species are complete competitors. Individuals are immobile in lattice sites and populations waves spread due to reproduction of individuals (Movies S1-S4). The closest biological analogue is vegetative reproduction of plants (Figs 3 and S1).

In order to model interspecific competition, we mechanistically define dominance as a primary ability of an individual of a species with greater fitness to occupy a free microhabitat in a direct conflict of interest with an individual of a less adapted species (Fig. 3). Rules of competitive interactions between species are represented in a general form by diagrams (Fig. 4).

These cellular automata models are individual-based and their rules consist of deterministic logical ‘if-then’ statements only (Figs 5 and S2). Thus, the models are fully mechanistic. The models take into account a regeneration state of a microhabitat after an individual's death. Earlier regenerative processes in ecosystems were considered in the regeneration niche concept[35,36]. For example, obstacles in the form of dead roots, dead tillers as and soil toxins must be eliminated and necessary mineral components must be recovered after propagating of population wave. Thus a set of conditions must be restored for a possibility of subsequent successful occupation of a microhabitat by an individual.

Inclusion of the regeneration state of a microhabitat allow us to implement the accordance of our models with the axiomatic formalism of Wiener and Rosenblueth for simulation of excitation propagation in active media[37]. Three successive states - rest, excitation and refractoriness of each site are the main features of that formalism. In our formalism the ‘rest’ state corresponds to the ‘free’ state of a microhabitat, the ‘excitation’ corresponds to the life activity of an individual in a microhabitat and the ‘refractoriness’ corresponds to the regeneration state of a microhabitat (Figs 5b-d and S2).

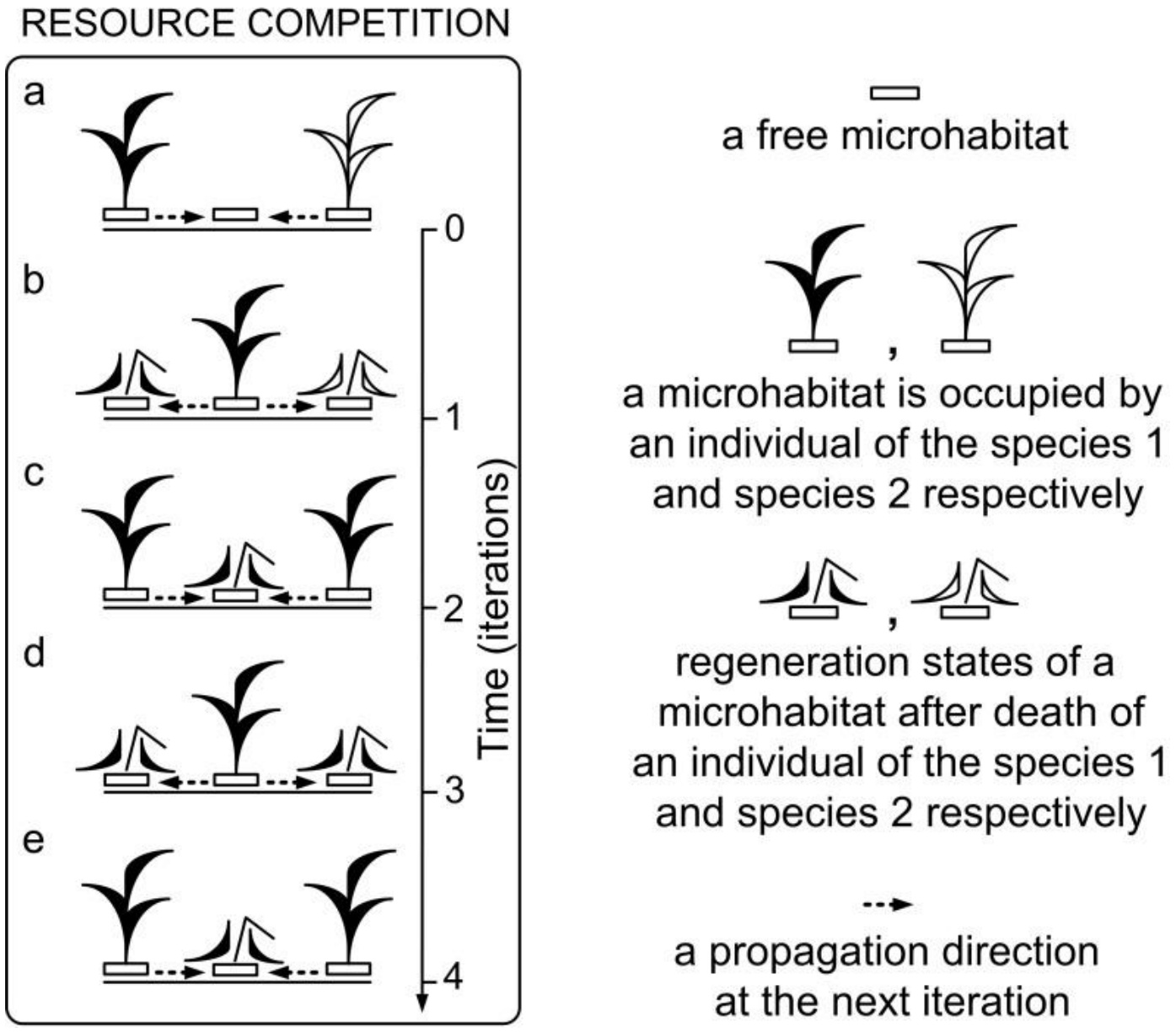

**Figure 3 | Interspecific competition.** The species 1 is dominant, i.e. it has greater fitness than the species 2. **a**, Individuals of competing species try to propagate into the same microhabitat. **b**, The better adapted species 1 wins the species 2. Dead individuals are recycled and spent resources are recovered during the regeneration state of a microhabitat. **c**-**e**, The first species has won and continues to live and propagate.

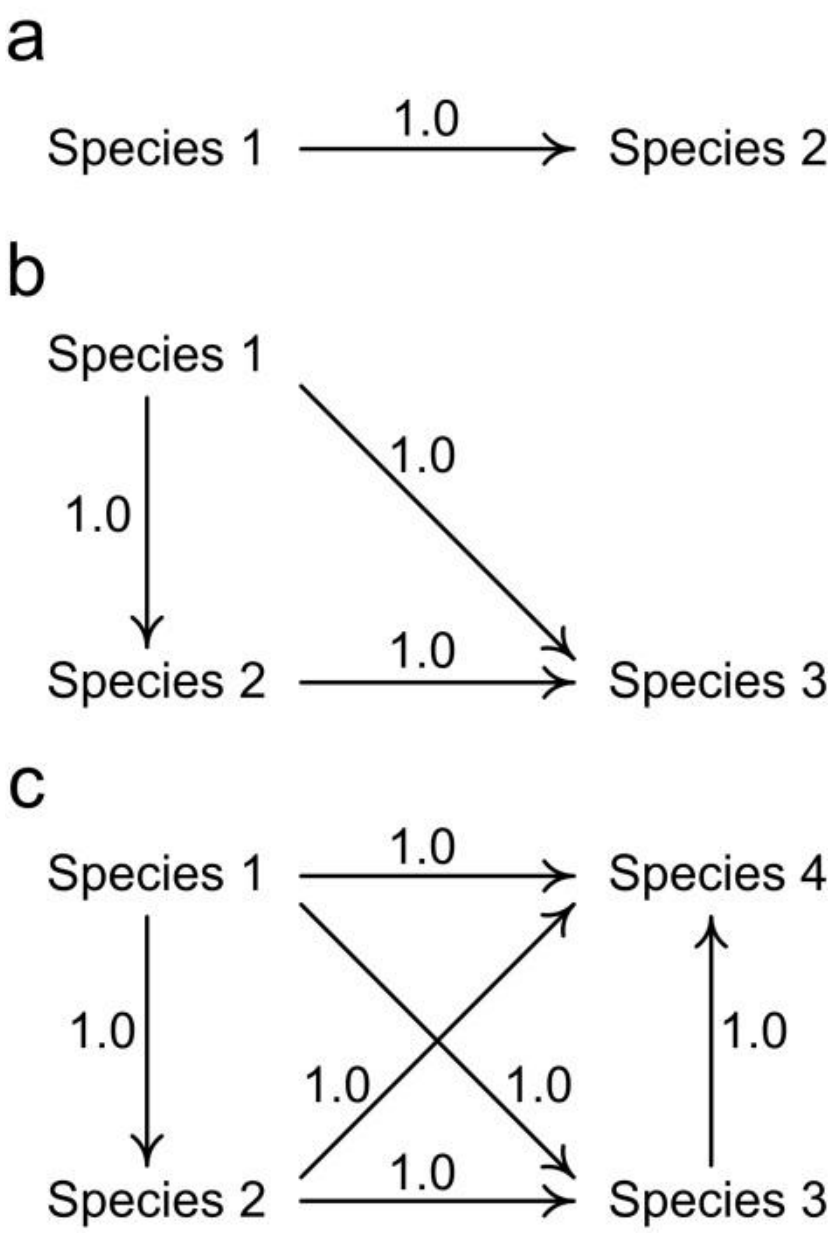

**Figure 4 | Diagrams of interspecific competitive interactions.** Arrows point from the more fit species to the less fit ones. The probability of occupation of a microhabitat in the free state by an offspring of an individual of the species with greater fitness in a direct conflict of interest is equal to 1.0, i.e. a dominant species wins in a direct conflict of interest in 100% of cases. **a**, A diagram of the two-species competition model. The species 1 wins the species 2. **b**, A diagram of the three-species competition model. The species 1 wins the species 2 and 3, while the species 2 wins the species 3. **c**, A diagram of the four-species competition model. The species 1 wins the species 2, 3 and 4. The species 2 wins the species 3 and 4. The species 3 wins the species 4 only.

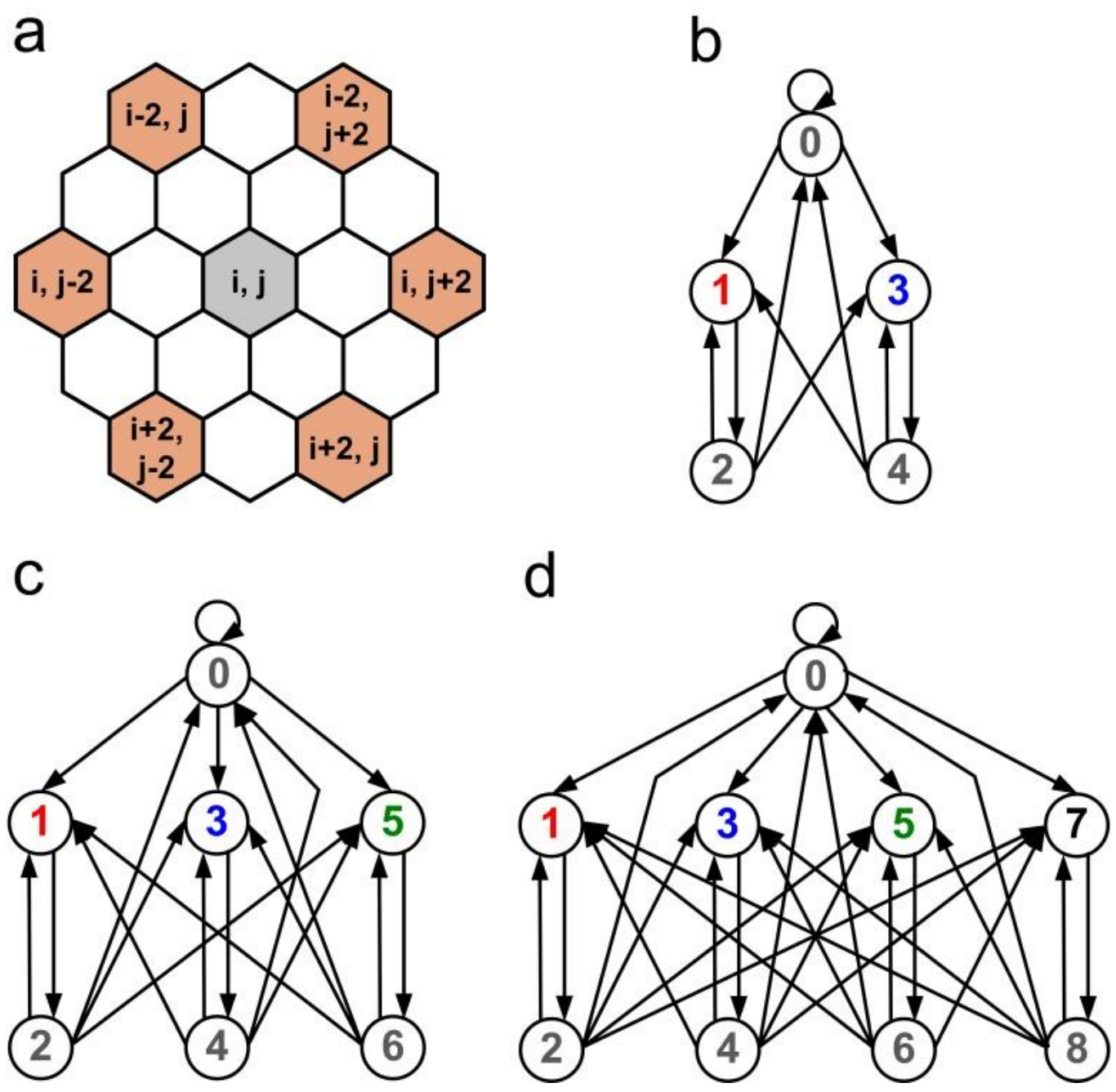

**Figure 5 | Rules of the cellular automata models. a**, A hexagonal rosette-like neighbourhood where i and j are integer numbers. A site with parental individual has coordinates (i, j) and is marked by the grey colour. Sites with possible offsprings of an individual of any species have coordinates (i, j-2), (i-2, j), (i-2, j+2), (i, j+2), (i+2, j), (i+2, j-2) and marked by the orange colour. **b-d**, Directed graphs of transitions between the states of a lattice site: in the two-species competition model (b), in the three-species competition model (c), in the four-species competition model (d). States of a lattice site are denoted as: 0 – a free microhabitat; 1, 3, 5, 7 – are the states of a microhabitat occupied by an individual of the first, second, third and fourth species, respectively. In Supplementary movies these states are marked by the same colours, but represented as symbols '1', '2', '3', '4' according to the number of the species; 2, 4, 6, 8 – are the regeneration states of a site after living of an individual of the first, second, third and fourth species respectively. In Supplementary movies these regeneration states are represented by symbols '.', '*', '@', '#', respectively, to distinguish them from living individuals.

A populated microhabitat goes into the regeneration state after an individual's death. A populated microhabitat and a microhabitat in the regeneration state cannot be occupied. A microhabitat can be occupied if it is in the free state or immediately after finishing of the regeneration state when it becomes free. Thus, we simulate a birth-death-regeneration process. In addition, the regeneration state of a microhabitat allows us to avoid a predator-prey analogy when an individual directly replaces another one. Logical rules of the two-species competition model are considered in Figure S2 in more details.

Population dynamics in the models with two, three and four competitors was investigated by Monte Carlo simulations with 200 repeated experiments of random initial placements of single individuals of competing species on the lattice consisting of 26x26 sites (Fig. 2).

Along with investigation of random initial positioning of individuals on the lattice consisted of 26x26 sites, we investigated population dynamics in the two-species competition model by the same Monte Carlo method with the following lattice sizes (NxN lattice sites): 27x27, 28x28, 29x29, 30x30 (Fig. S3) and, 100x100, 101x101, 102x102, 103x103 (Fig. S4).

Additionally, to analyse population dynamics we performed 100 repeated series of Monte Carlo simulations with 200 repeated experiments in every Monte Carlo simulation (Fig. S5-S8). Random initial placements of single individuals of competing species on the lattice were used in each of 200 repeated experiments (Monte Carlo simulation). The lattice consists of 26x26 sites.

**Acknowledgements**

We thank A. S. Komarov for inspiring our interest in ecological modelling. We thank A. V. Kalmykov and A. E. Noble for helpful discussions. We thank M. Fujiwara for his valuable comment.

**Author Contributions**

V.L.K. designed the research. L.V.K. created the programs, investigated the models and made the figures and the movies of the experiments. Both authors discussed the results, interpreted them and wrote the manuscript.

**Additional information**

Competing financial interests: The authors declare no competing financial interests.

## References

1 Craze, P. *Ecological neutral theory: useful model or statement of ignorance? Cell Press Didscussions*, URL:http://news.cell.com/discussions/trends-in-ecology-and-evolution/ecological-neutral-theory-useful-model-or-statement-of-ignorance (2012).

2 Clark, J. S. The coherence problem with the Unified Neutral Theory of Biodiversity. *Trends in Ecology & Evolution* **27**, 198-202, doi:10.1016/j.tree.2012.02.001 (2012).

3 Palmer, M. W. Variation in Species Richness - Towards a Unification of Hypotheses. *Folia Geobot Phytotx* **29**, 511-530, doi:10.1007/BF02883148 (1994).

4 Gause, G. F. *The struggle for existence.* (NY: Hafner Publishing Company, 1934).

5 Hardin, G. The Competitive Exclusion Principle. *Science* **131**, 1292-1297, doi:10.1126/science.131.3409.1292 (1960).

6 Hutchinson, G. E. The paradox of the plankton. *The American Naturalist* **95**, 137-145, doi:10.2307/2458386 (1961).

7 Kerr, B., Riley, M. A., Feldman, M. W. & Bohannan, B. J. M. Local dispersal promotes biodiversity in a real-life game of rock-paper-scissors. *Nature* **418**, 171-174, doi:10.1038/nature00823 (2002).

8 Vandermeer, J. & Yitbarek, S. Self-organized spatial pattern determines biodiversity in spatial competition. *Journal of Theoretical Biology* **300**, 48-56, doi:10.1016/j.jtbi.2012.01.005 (2012).

9 Reichenbach, T., Mobilia, M. & Frey, E. Mobility promotes and jeopardizes biodiversity in rock-paper-scissors games. *Nature* **448**, 1046-1049, doi:10.1038/nature06095 (2007).

10 Laird, R. A. & Schamp, B. S. Competitive intransitivity promotes species coexistence. *American Naturalist* **168**, 182-193, doi:10.1086/506259 (2006).

11 Allesina, S. & Levine, J. M. A competitive network theory of species diversity. *Proceedings of the National Academy of Sciences* **108**, 5638-5642, doi:10.1073/pnas.1014428108 (2011).

12 Tilman, D. The importance of the mechanisms of interspecific competition. *The American Naturalist* **129**, 769-774, doi:10.1086/284672 (1987).

13 Ivanitsky, G. R., Medvinsky, A. B. & Tsyganov, M. A. From the dynamics of population autowaves generated by living cells to neuroinformatics. *Physics-Uspekhi* **37**, 961-989, doi:dx.doi.org/10.1070/PU1994v037n10ABEH000049 (1994).

14 Krinsky, V. I. Autowaves: Results, problems, outlooks in *Self-Organization Autowaves and Structures Far from Equilibrium* Vol. 28 Springer Series in Synergetics (ed V. I. Krinsky) Ch. 2, 9-19 (Springer Berlin Heidelberg, 1984). doi:10.1007/978-3-642-70210-5_2.

15 Zaikin, A. N. & Zhabotinsky, A. M. Concentration Wave Propagation in Two-dimensional Liquid-phase Self-oscillating System. *Nature* **225**, 535-537, doi:10.1038/225535b0 (1970).

16 Winfree, A. T. Persistent tangled vortex rings in generic excitable media. *Nature* **371**, 233-236, doi:10.1038/371233a0 (1994).

17 Tsyganov, M. A., Kresteva, I. B., Medvinskii, A. B. & Ivanitskii, G. R. A Novel Mode of Bacterial Population Wave Interaction. *Doklady Akademii Nauk (In Russian)* **333**, 532-536, Accession number ISI:A1993MT05900031 (1993).

18 Tsyganov, M. & Ivanitsky, G. Solitonlike and nonsoliton modes of interaction of taxis waves (illustrated with an example of bacterial population waves). *Biophysics* **51**, 887-891, doi:10.1134/S0006350906060066 (2006).

19 Tsyganov, M. A., Biktashev, V. N., Brindley, J., Holden, A. V. & Ivanitsky, G. R. Waves in systems with cross-diffusion as a new class of nonlinear waves. *Physics-Uspekhi* **50**, 263-286, doi:10.1070/PU2007v050n03ABEH006114 (2007).

20 Borner, U., Deutsch, A., Reichenbach, H. & Bar, M. Rippling patterns in aggregates of myxobacteria arise from cell-cell collisions. *Physical review letters* **89**, 078101, doi:10.1103/PhysRevLett.89.078101 (2002).

21 Igoshin, O. A., Mogilner, A., Welch, R. D., Kaiser, D. & Oster, G. Pattern formation and traveling waves in myxobacteria: theory and modeling. *Proceedings of the National Academy of Sciences of the United States of America* **98**, 14913-14918, doi:10.1073/pnas.221579598 (2001).

22 Igoshin, O. A. & Oster, G. Rippling of myxobacteria. *Mathematical biosciences* **188**, 221-233, doi:10.1016/j.mbs.2003.04.001 (2004).

23 Sliusarenko, O., Neu, J., Zusman, D. R. & Oster, G. Accordion waves in Myxococcus xanthus. *Proc Natl Acad Sci USA* **103**, 1534-1539, doi:10.1073/pnas.0507720103 (2006).

24 Welch, R. & Kaiser, D. Cell behavior in traveling wave patterns of myxobacteria. *Proceedings of the National Academy of Sciences* **98**, 14907-14912, doi:10.1073/pnas.261574598 (2001).

25 Sliusarenko, O., Chen, J. & Oster, G. From biochemistry to morphogenesis in myxobacteria. *Bulletin of mathematical biology* **68**, 1039-1051, doi:10.1007/s11538-006-9113-9 (2006).

26 Kalmykov, L. V. & Kalmykov, V. L. Mechanistic mechanisms of competition and biodiversity. Available from Nature Precedings doi:hdl.handle.net/10101/npre.2012.7105.1 (2012).

27 Estrada-Flores, S., Merts, I., De Ketelaere, B. & Lammertyn, J. Development and validation of “grey-box” models for refrigeration applications: A review of key concepts. *International Journal of Refrigeration* **29**, 931-946, doi:10.1016/j.ijrefrig.2006.03.018 (2006).

28 Kroll, A. *Grey-box models: Concepts and application. In: New Frontiers in Computational Intelligence and its Applications, vol.57 of Frontiers in artificial intelligence and applications*. 42-51 (IOS Press, 2000).

29 Engel, A. *Verification, validation, and testing of engineered systems*. (Wiley, 2010).

30 Wilson, J. B. Mechanisms of Species Coexistence - 12 Explanations for Hutchinson Paradox of the Plankton - Evidence from New-Zealand Plant-Communities. *New Zealand Journal of Ecology* **13**, 17-42, doi:10.1111/j.1654-1103.2010.01226.x (1990).

31 Sommer, U. Ecology - Competition and coexistence. *Nature* **402**, 366-367, doi:10.1038/46453 (1999).

32 Macarthur, R. & Levins, R. Competition, habitat selection, and character displacement in a patchy environment. *Proceedings of the National Academy of Sciences of the United States of America* **51**, 1207-1210, doi:10.1073/pnas.51.6.1207 (1964).

33 Levin, S. A. Community Equilibria and Stability, and an Extension of the Competitive Exclusion Principle. *Am Nat* **104**, 413-423, URL: http://www.jstor.org/stable/2459310 (1970).

34 Darlington, P. J., Jr. Competition, competitive repulsion, and coexistence. *Proceedings of the National Academy of Sciences of the United States of America* **69**, 3151-3155, URL:http://www.pnas.org/content/69/11/3151.full.pdf+html (1972).

35 Watt, A. S. Pattern and Process in the Plant Community. *Journal of Ecology* **35**, 1-22, doi:10.2307/2256497 (1947).

36 Grubb, P. J. The maintenance of species-richness in plant communities: the importance of the regeneration niche. *Biological Reviews* **52**, 107-145, doi:10.1111/j.1469-185X.1977.tb01347.x (1977).

37 Wiener, N. & Rosenblueth, A. The mathematical formulation of the problem of conduction of impulses in a network of connected excitable elements, specifically in cardiac muscle. *Archivos del Instituto de Cardiologia de Mexico* **16**, 205-265 (1946).

# Supplementary Information

# A mechanistic verification of the competitive exclusion principle

Lev V. Kalmykov[1,3] and Vyacheslav L. Kalmykov[2,3]*

[1]*Institute of Theoretical and Experimental Biophysics, Russian Academy of Sciences, Pushchino, Moscow Region, 142290 Russia*

[2]*Institute of Cell Biophysics, Russian Academy of Sciences, Pushchino, Moscow Region, 142290 Russia*

[3]*Pushchino State Institute of Natural Sciences (the former Pushchino State University), Pushchino, Moscow Region, 142290 Russia*

* e-mail: vyacheslav.l.kalmykov@gmail.com

## Contents

### Supplementary figures

### Supplementary movies (legends)

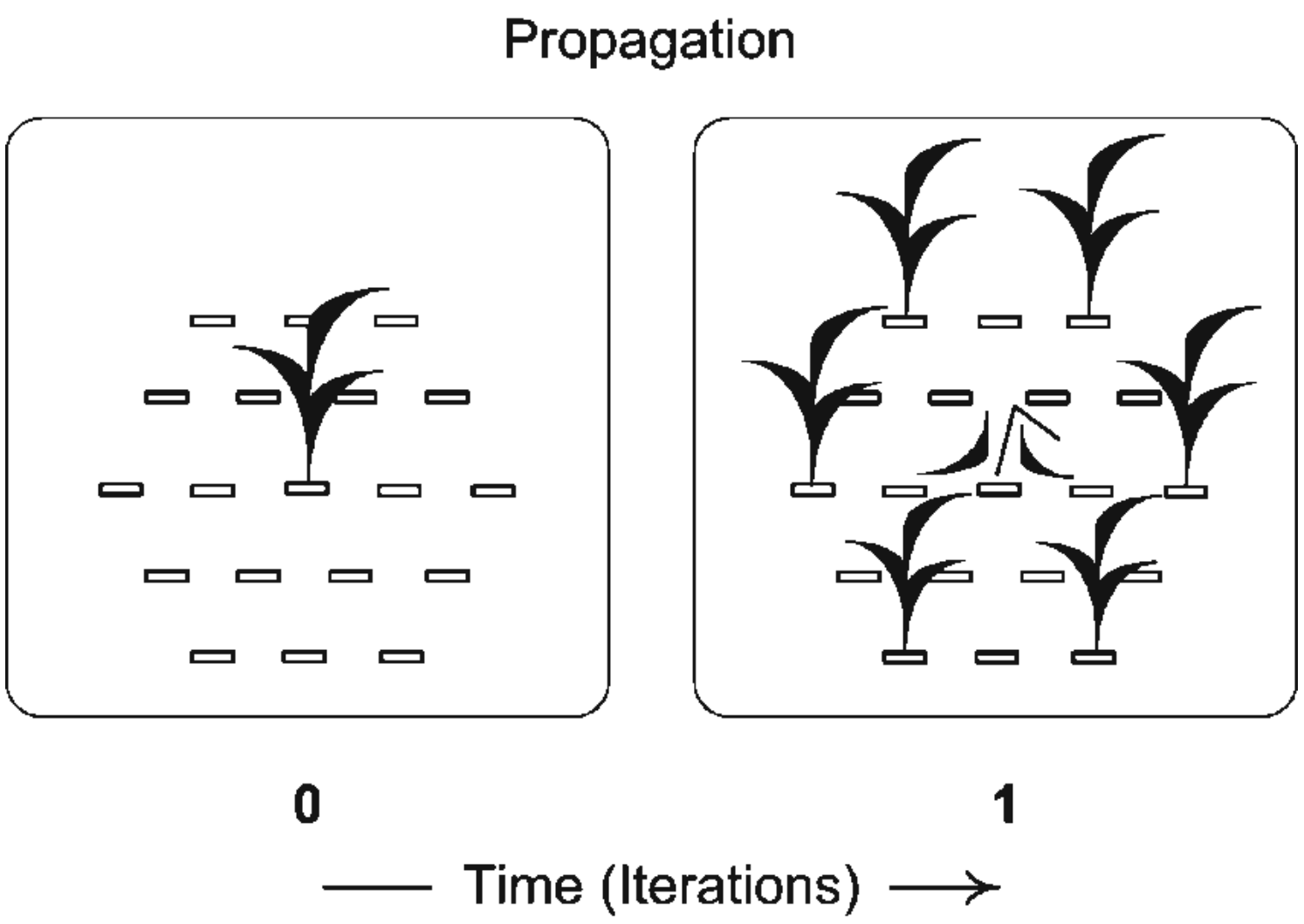

**Figure S1 | Moderate vegetative reproduction of plants.** A pattern of propagation and number of offsprings are determined by the hexagonal rosette-like neighbourhood.

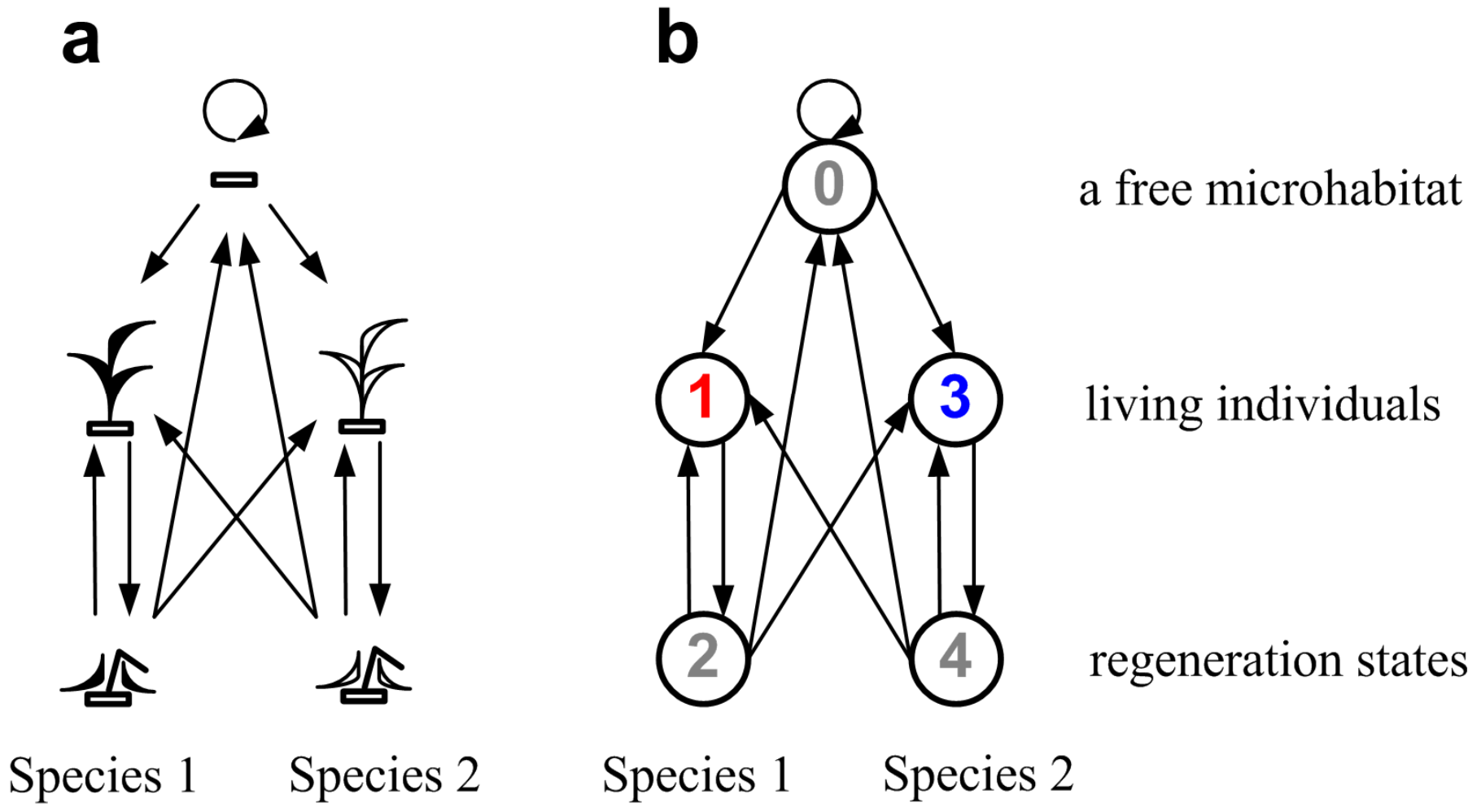

**Figure S2 | Rules of the two-species competition model.** A graph of transitions between the states of a site (microhabitat) in the two-species competition model is represented here in two forms (a and b). **a**, Pictorial form. **b**, Numerical form of program implementation.

Each microhabitat may be in one of the five states:

0 – a free microhabitat that can be occupied by an offspring of any species;

1 – a microhabitat which is occupied by a living individual of the first species;

2 – a regeneration state of a microhabitat after death of an individual of the first species;

3 – a microhabitat which is occupied by a living individual of the second species;

4 – a regeneration state of a microhabitat after death of an individual of the second species.

Logical rules of transitions between the states of a microhabitat in the two-species competition model:

0→0, a microhabitat remains free if in its neighbourhood is no one living individual;

0→1, a microhabitat will be occupied by an individual of the first species if in its neighbourhood is at least one individual of the first species;

0→3, a microhabitat will be occupied by an individual of the second species if in its neighbourhood is at least one individual of the second species and there is no one individual of the first species;

1→2, after death of an individual of the first species its microhabitat goes into the regeneration state;

2→0, after the regeneration state a microhabitat will become free if in its neighbourhood is no one living individual;

2→1, after the regeneration state a microhabitat will be occupied by an individual of the first species if in its neighbourhood is at least one individual of the first species;

2→3, after the regeneration state a microhabitat will be occupied by an individual of the second species if in its neighbourhood is at least one individual of the second species and there is no one individual of the first species;

3→4, after death of an individual of the second species its microhabitat goes into the regeneration state;

4→0, after the regeneration state a microhabitat will become free if in its neighbourhood is no one living individual;

4→1, after the regeneration state a microhabitat will be occupied by an individual of the first species if in its neighbourhood is at least one individual of the first species;

4→3, after the regeneration state a microhabitat will be occupied by an individual of the second species if in its neighbourhood is at least one individual of the second species and there is no one individual of the first species.

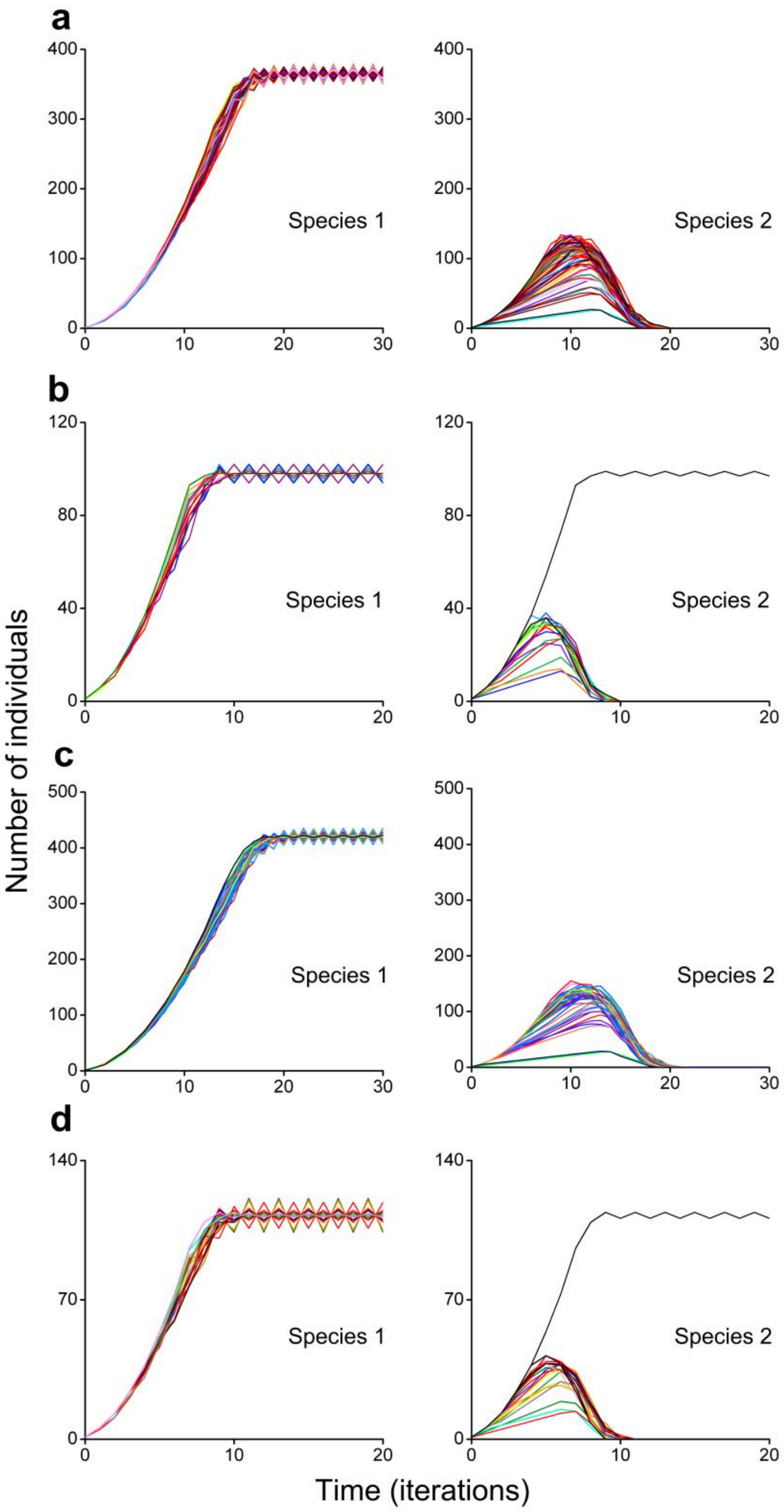

**Figure S3 | Population dynamics in the two-species competition model with small lattice sizes.** Results of the Monte Carlo simulation are presented here. The Monte Carlo simulation consists of 200 repeated experiments with random placements of single individuals of the two competing species on the lattice at the initial iteration of the cellular automata model.

**a**, The lattice consists of 27x27 sites. Species 1 excludes species 2.

**b**, The lattice consists of 28x28 sites. Cases of coexistence and exclusion.

**c**, The lattice consists of 29x29 sites. Species 1 excludes species 2.

**d**, The lattice consists of 30x30 sites. Cases of coexistence and exclusion.

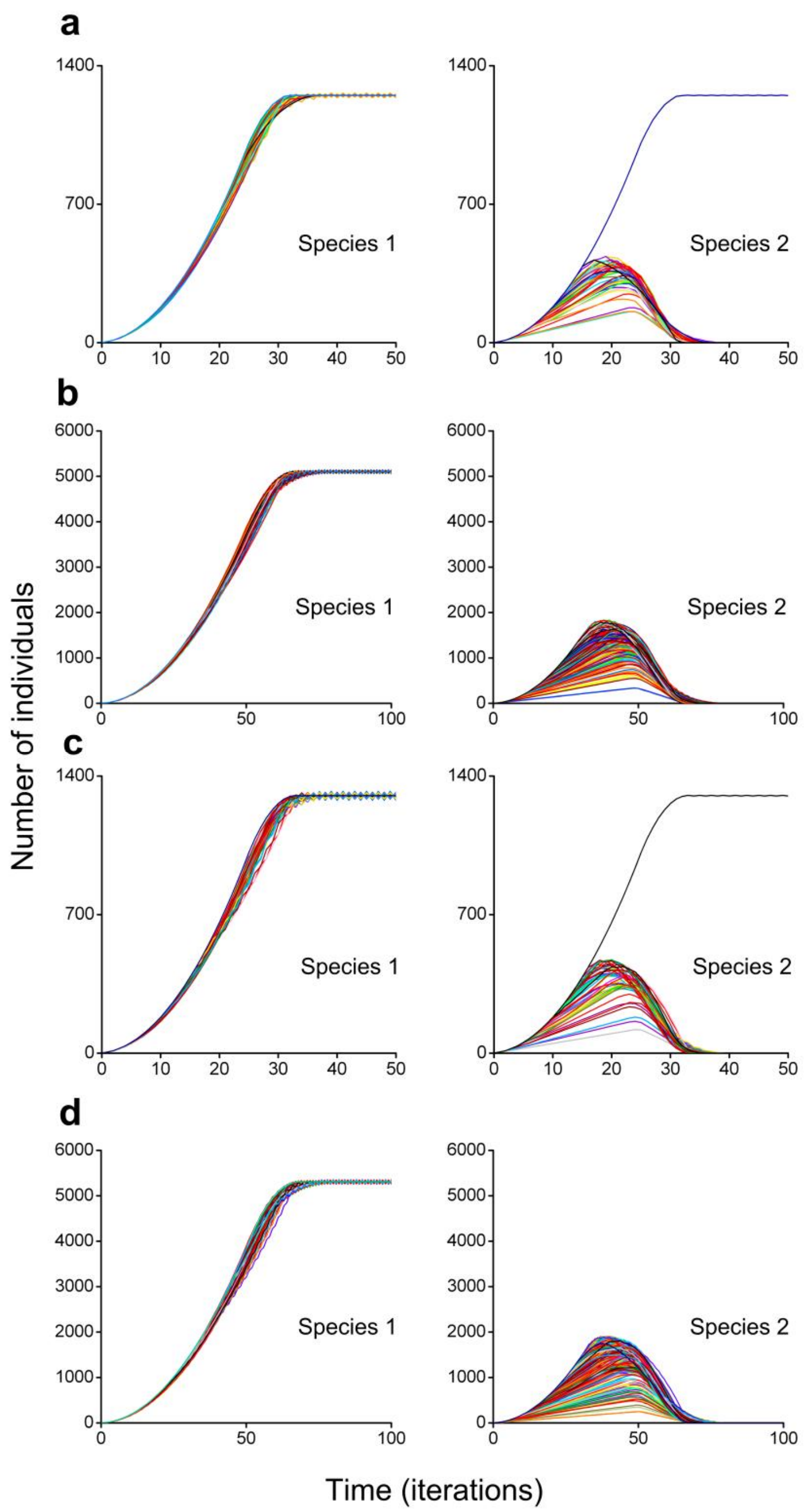

**Figure S4 | Population dynamics in the two-species competition model with large lattice sizes.** Results of the Monte Carlo simulation are presented here. The Monte Carlo simulation consists of 200 repeated experiments with random placements of single individuals of the two competing species on the lattice at the initial iteration of the cellular automata model.

**a,** The lattice consists of 100x100 sites. Cases of coexistence and exclusion.

**b**, The lattice consists of 101x101 sites. Species 1 excludes species 2.

**c**, The lattice consists of 102x102 sites. Cases of coexistence and exclusion.

**d**, The lattice consists of 103x103 sites. Species 1 excludes species 2.

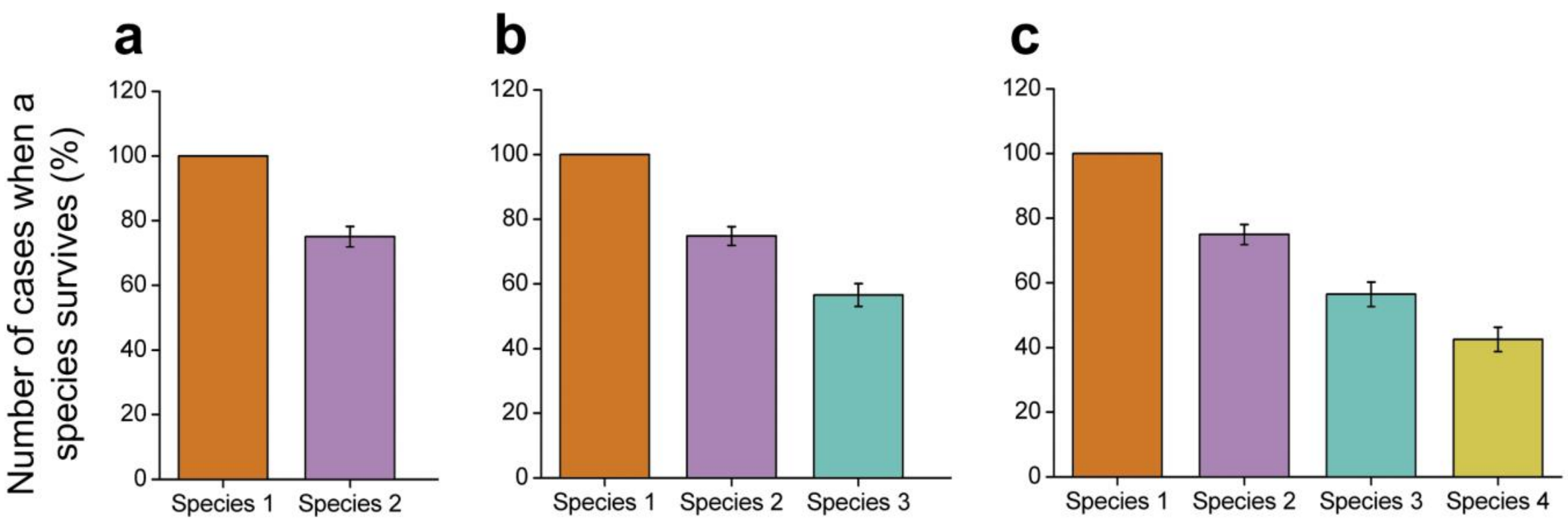

**Figure S5 | Survival of the species in the models with two, three and four competitors.** The number of cases when a species survives (mean ± 1 SD; n=100 *series*). Each of these 100 repeated *series* consists of 200 repeated Monte Carlo experiments with random placements of single individuals of the competing species on the lattice at the initial iteration of the cellular automata model. Cellular automata lattice consists of 26x26 sites.

**a**, The two-species competition model.

**b**, The three-species competition model.

**c**, The four-species competition model.

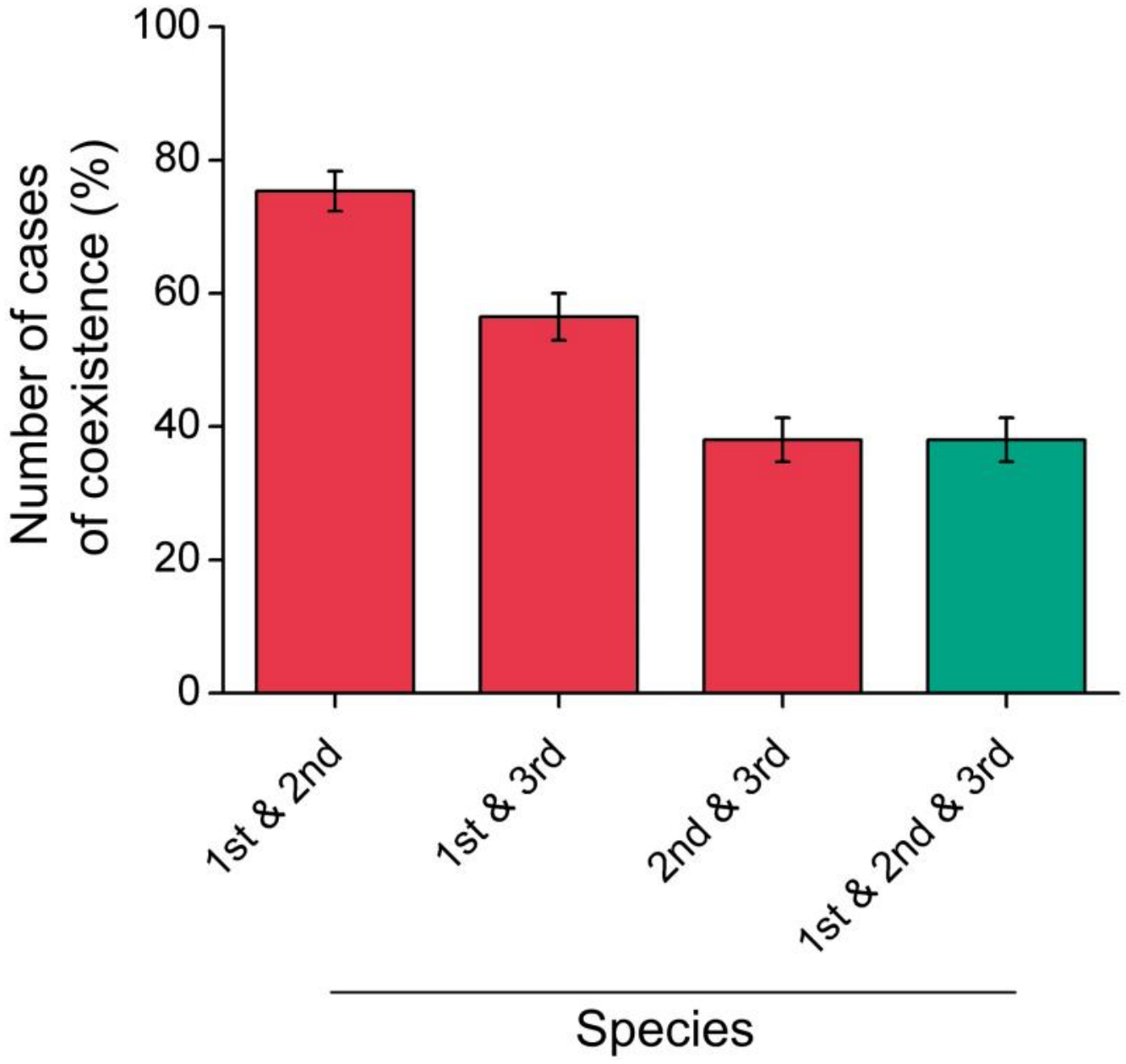

**Figure S6 | Coexistence of the species in the three-species competition model.** The number of cases of coexistence of the species (mean ± 1 SD; n=100 *series*). Each of these 100 repeated *series* consists of 200 repeated Monte Carlo experiments with random placements of single individuals of the competing species on the lattice at the initial iteration of the cellular automata model. Cellular automata lattice consists of 26x26 sites.

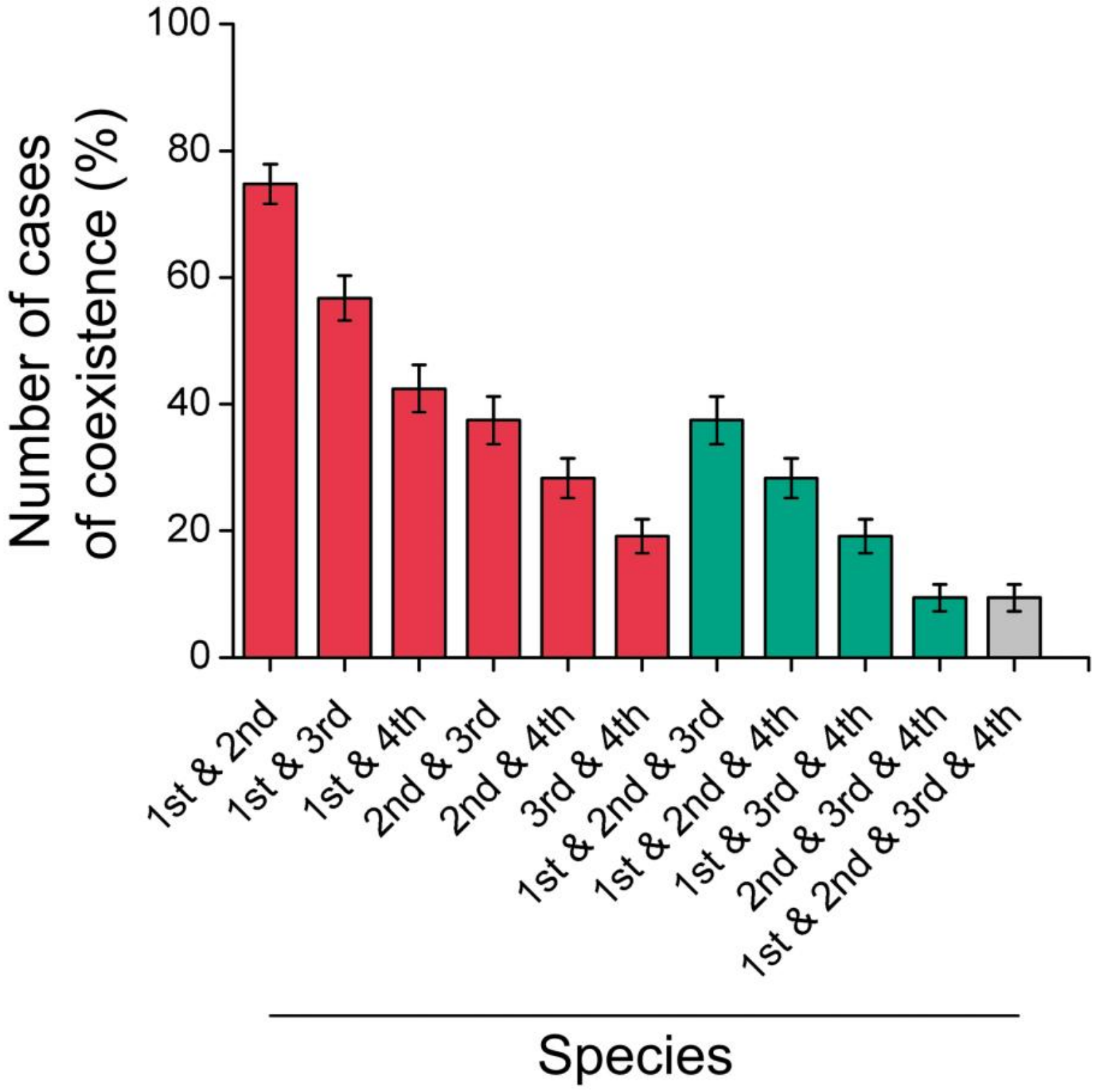

**Figure S7 | Coexistence of the species in the four-species competition model.**

The number of cases of coexistence of the species (mean ± 1 SD; n=100 *series*). Each of these 100 repeated *series* consists of 200 repeated Monte Carlo experiments with random placements of single individuals of the competing species on the lattice at the initial iteration of the cellular automata model. Cellular automata lattice consists of 26x26 sites.

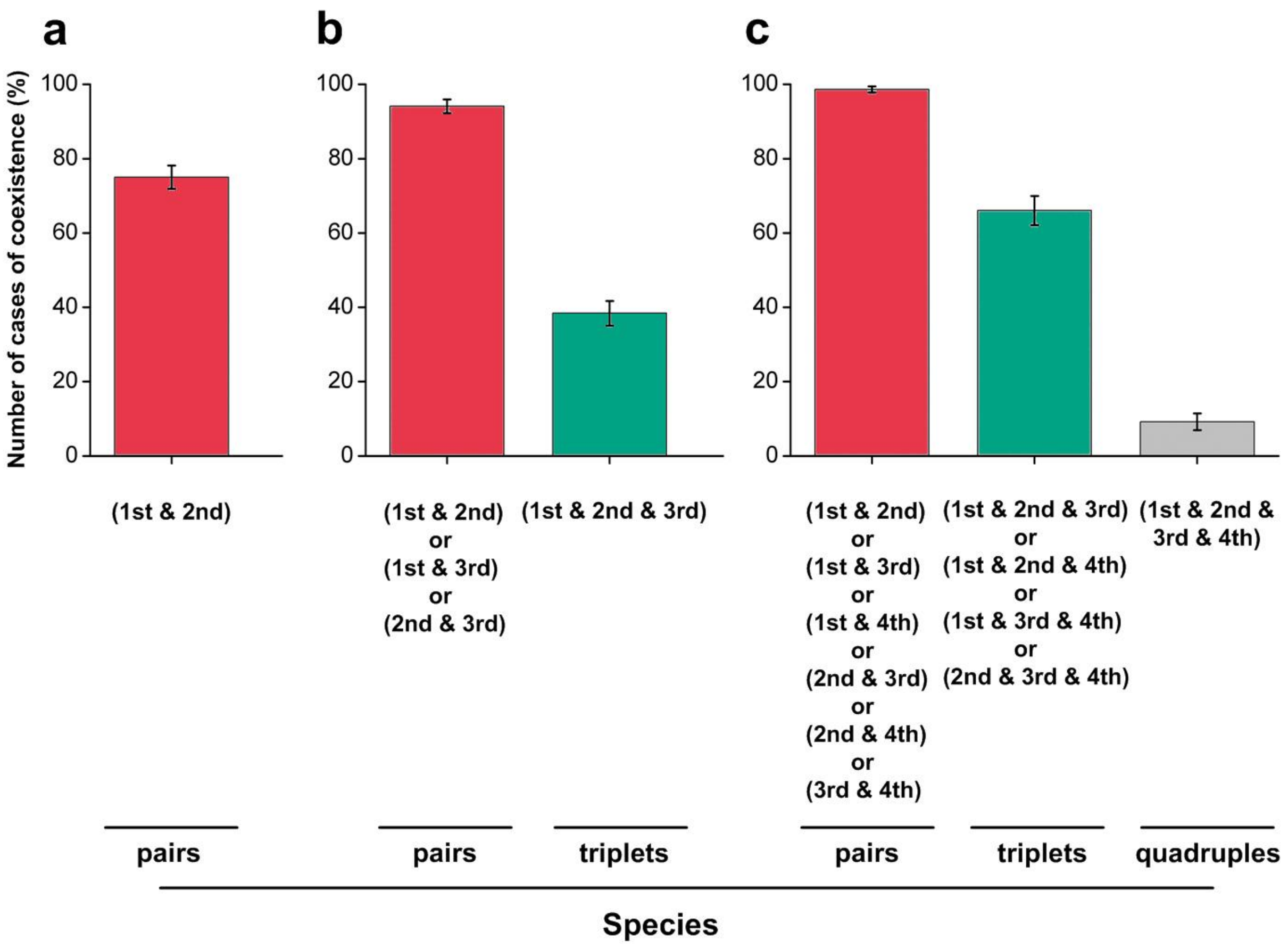

**Figure S8 | Coexistence of the species in the models with two, three and four competitors.**

The number of cases of coexistence of the competing species (mean ± 1 SD; n=100 *series*). Each of these 100 repeated *series* consists of 200 repeated Monte Carlo experiments with random placements of single individuals of the competing species on the lattice at the initial iteration of the cellular automata model. Cellular automata lattice consists of 26x26 sites.

**a**, The two-species competition model. The number of cases of coexistence is 150.06 ± 6.26, i.e. 75.03% ± 3.13% of cases.

**b**, The three-species competition model. The numbers of cases of coexistence are: 76.8 ± 6.65 (38.4% ± 3.33%) for the three competing species and 188.17 ± 3.67 (94.09% ± 1.84%) for pairs of competing species, i.e. it is the number of cases when there is at least one pair of coexisting competing species.

**c**, The four-species competition model. The numbers of cases of coexistence are: 18.44 ± 4.52 (9.22% ± 2.26%) for the four competing species, 132.06 ± 7.79 (66.03% ± 3.9%) for triplets of competing species, i.e. it is the number of cases when there is at least one triplet of coexisting competing species, and 197.28 ± 1.69 (98.64% ± 0.85%) for pairs of competing species, i.e. it is the number of cases when there is at least one pair of coexisting competing species.

## Supplementary Movies

**Movie S1 | Coexistence of two competing species.** A logical deterministic individual-based cellular automata model of two-species competition. Two species coexist due to a soliton-like interpenetration of colliding population waves.

**Movie S2 | A case of competitive exclusion.** A logical deterministic individual-based cellular automata model of two-species competition.

**Movie S3 | Coexistence of three competing species.** A logical deterministic individual-based cellular automata model of three-species competition. Three species coexist due to a soliton-like interpenetration of colliding population waves.

**Movie S4 | Coexistence of four competing species.** A logical deterministic individual-based cellular automata model of four-species competition. Four species coexist due to a soliton-like interpenetration of colliding population waves.